\documentclass{ws-procs9x6}

\def\citebk#1{[\hspace{0.3mm}\raisebox{-1.85mm}[0mm][0mm]
  {\Large\cite{#1}}\hspace{-0.1mm}]}

\begin{document}

\title{Quark color superconductivity and the cooling of compact stars}

\author{I.~A. SHOVKOVY AND P.~J. ELLIS}

\address{School of Physics and Astronomy,\\
University of Minnesota, \\
Minneapolis, MN 55455, USA}

\maketitle

\abstracts{
The thermal conductivity of the color-flavor locked phase of dense quark
matter is calculated. The dominant contribution to the conductivity comes
from photons and Nambu-Goldstone bosons associated with the breaking of
baryon number, both of which are trapped in the quark core. Because of
their very large mean free path the conductivity is also very large. The
cooling of the quark core arises mostly from the heat flux across the
surface of direct contact with the nuclear matter. As the thermal
conductivity of the neighboring layer is also high, the whole interior of
the star should be nearly isothermal. Our results imply that the cooling
time of compact stars with color-flavor locked quark cores is similar to
that of ordinary neutron stars.}

\section{Introduction}

At sufficiently high baryon density the nucleons in nuclear matter should
melt into quarks so that the system becomes a quark liquid. It should be
weakly interacting due to asymptotic freedom \citebk{ColPer}, however, it
cannot be described as a simple Fermi liquid. This is due to the
nonvanishing attractive interaction in the color antitriplet quark-quark
channel, provided by one-gluon exchange, which renders a highly
degenerate Fermi surface unstable with respect to Cooper pairing.  As a
result the true ground state of dense quark matter is, in fact, a color
superconductor \citebk{old}.

Recent phenomenological \citebk{W1S1} and microscopic studies
\citebk{PR1-Son,2nd-wave,3rd-wave,us2} have confirmed that quark matter
at a sufficiently high density undergoes a phase transition into a color
superconducting state. Phenomenological studies are expected to be
appropriate to intermediate baryon densities, while microscopic
approaches are strictly applicable at asymptotic densities where
perturbation theory can be used. It is remarkable that both approaches
concur that the superconducting order parameter (which determines the gap
$\Delta$ in the quark spectrum) lies between $10$ and $100$ MeV for
baryon densities existing in the cores of compact stars.

At realistic baryon densities only the three lightest quarks can
participate in the pairing dynamics.  The masses of the quarks are much
smaller than the baryon chemical potential, thus, to a good
approximation, all three flavors participate equally in the color
condensation. The ground state is then the so-called color flavor locked
(CFL) phase \citebk{ARW}. The original gauge symmetry $SU(3)_{c}$ and the
global chiral symmetry $SU(3)_{L} \times SU(3)_{R}$ break down to a
global diagonal ``locked" $SU(3)_{c+L+R}$ subgroup. Because of the Higgs
mechanism the gluons become massive and decouple from the infrared
dynamics. The quarks also decouple because large gaps develop in their
energy spectra. The breaking of the chiral symmetry leads to the
appearance of an octet of pseudo-Nambu-Goldstone (NG) bosons ($\pi^{0}$,
$\pi^{\pm}$, $K^{\pm}$, $K^{0}$, $\bar{K}^{0}$, $\eta$). In addition an
extra NG boson $\phi$ and a pseudo-NG boson $\eta^{\prime}$ appear in the
low energy spectrum as a result of the breaking of global baryon number
symmetry and approximate $U(1)_{A}$ symmetry, respectively.

The low energy action for the NG bosons in the limit of asymptotically
large densities was derived in Refs.~\citebk{CasGat,SonSt}. By making use
of an auxiliary ``gauge" symmetry, it was suggested in Ref.~\citebk{BS}
that the low energy action of Refs.~\citebk{CasGat,SonSt} should be
modified by adding a time-like covariant derivative to the action of the
composite field. Under a favorable choice of parameters, the modified
action predicted kaon condensation in the CFL phase. Some unusual
properties of such a condensate were discussed in Ref.~\citebk{MirSho}.

While the general structure of the low energy action in the CFL phase can
be established by symmetry arguments alone \citebk{CasGat}, the values of
the parameters in such an action can be rigorously derived only at
asymptotically large baryon densities \citebk{SonSt,BS}. Thus, in the
most interesting case of intermediate densities existing in the cores of
compact stars, the details of the action are not well known. For the
purposes of the present paper, however, it suffices to know that there
are 9 massive pseudo-NG bosons and one massless NG boson $\phi$ in the
low energy spectrum. If kaons condense \citebk{BS} an additional NG boson
should appear. These NG bosons should be relevant for the kinetic
properties of dense quark matter.

It has been found \citebk{JPS} that neutrino and photon emission rates
for the CFL phase are very small so that they would be inefficient in
cooling the core of a neutron star. The purpose of the present
investigation is to determine quantitatively the thermal conductivity of
the CFL phase of dense quark matter in order to see whether it can
significantly impact the cooling rate.  We shall argue that the
temperature of the CFL core, as well as the neighboring neutron layer
which is in contact with the core, falls quickly due to the very high
thermal conductivities on both sides of the interface. In fact, to a good
approximation, the interior of the star is isothermal. A noticeable
gradient of the temperature appears only in a relatively thin surface
layer of the star where a finite flux of energy is carried outwards by
photon diffusion. A more complete account of this work is to be found in
Ref.~\citebk{SE-prc}.

\section{Thermal conductivity}
\label{thcond}

A detailed understanding of the cooling mechanism of a compact star with
a quark core is not complete without a study of thermal conductivity
effects in the color superconducting quark core. The conductivity, as
well as the other kinetic properties of quark matter in the CFL phase, is
dominated by the low energy degrees of freedom. It is clear then that at
all temperatures of interest to us, $T\ll \Delta$, it is crucial to
consider the contributions of the NG bosons. In addition, there may be an
equally important contribution due to photons; this is discussed in
Sec.~\ref{photon}. Note that, at such small temperatures, the gluon and
the quark quasiparticles become completely irrelevant. For example, a
typical quark contribution to a transport coefficient would be
exponentially suppressed by the factor $\exp(-\Delta/T)$.

Let us start from the general definition of the thermal conductivity as a
characteristic of a system which is forced out of equilibrium by a
temperature gradient. In response to such a gradient transport of heat 
is induced. Formally this is described by the following relation:
\begin{equation}
u_{i} = - \kappa \partial_{i} T,
\end{equation}
where $u_{i}$ is the heat current, and $\kappa$ is the heat conductivity.
As is clear from this relation, the heat flow would persist until a state
of uniform temperature is reached. The higher the conductivity, the
shorter the time for this relaxation.

In the linear response approximation, the thermal conductivity is given
in terms of the heat current correlator by a Kubo-type formula.  We
derive the expression for the heat (energy) current carried by a single
(pseudo-) NG boson field $\varphi$. The corresponding Lagrangian density
reads
\begin{equation}
L = \frac{1}{2}\left(\partial_{0}\varphi \partial_{0}\varphi
- v^{2} \partial_{i} \varphi \partial_{i} \varphi 
- m^2 \varphi^{2} \right) +\ldots,
\end{equation}
where the ellipsis stand for the self-interaction terms as well as
interactions with other fields. Notice that we introduced explicitly the
velocity parameter $v$. In microscopic studies of color superconducting
phases, which are valid at very large densities, this velocity is equal
to $1/\sqrt{3}$ for all (pseudo-)  NG bosons. It is smaller than $1$
because Lorentz symmetry is broken due to the finite value of the quark
chemical potential. By making use of the above Lagrangian density, we
derive the following expression for the heat current:
\begin{equation}
u_{i} = \frac{\partial L}{\partial (\partial^{i} \varphi)}
\partial_{0} \varphi 
= v^{2} \partial_{i} \varphi \partial_{0} \varphi .
\label{heat-cur}
\end{equation}
This definition leads to an expression for the heat conductivity 
in terms of the corresponding correlator \citebk{fgi}:
\begin{equation}
\kappa_{ij} = -\frac{i}{2 T} \lim_{\Omega \to 0} \frac{1}{\Omega}
\left[\Pi_{ij}^{R}(\Omega+i\epsilon)-\Pi_{ij}^{A}(\Omega-i\epsilon)
\right],
\end{equation}
where, in the Matsubara formalism, 
\begin{eqnarray}
\Pi_{ij}(i\Omega_{m}) &=& v^{4} T \sum_{n} \int \frac{d^3 k}{(2\pi)^{3}}
k_{i} k_{j} i\Omega_{n} (i\Omega_{n} +i\Omega_{n-m}) \nonumber \\
&&\qquad\qquad\qquad\qquad\times S(i\Omega_{n}, \vec{k}) 
S(i\Omega_{n-m},\vec{k}) .
\end{eqnarray}
Here $\Omega_{n}\equiv 2\pi n T$ is the bosonic Matsubara frequency,
and $ S(i\Omega_{n}, \vec{k}) $ is the propagator of the (pseudo-) NG
boson. In general, the propagator should have the following form:
\begin{equation}
S(\omega,\vec{k}) = \frac{1}
{(\omega+i\Gamma/2)^{2}-v^{2}\vec{k}^{2}-m^2},
\label{propagator}
\end{equation}
where the width parameter $\Gamma(\omega, \vec{k})$ is related to the 
inverse lifetime (as well as the mean free path) of the boson.
In our calculation, it is very convenient to utilize the spectral
representation of the propagator,
\begin{equation}
S(i\Omega_{n},\vec{k}) =\frac{1}{\pi} \int_{-\infty}^{\infty}
\frac{d\omega A(\omega,\vec{k})}{i\Omega_{n}-\omega}.
\end{equation}
Then the conductivity is expressed through the spectral function
$A(\omega,\vec{k})$ as follows:
\begin{equation}
\kappa_{ij} = \frac{v^{4}}{2\pi T^2}  \int_{-\infty}^{\infty} 
\frac{\omega^2 d\omega }{\sinh^{2}\!\frac{\omega}{2T}}
\int \frac{d^3 k}{(2\pi)^{3}} k_{i} k_{j} 
A^{2}(\omega,\vec{k}).
\label{kap-general}
\end{equation}
By making use of the explicit form of the propagator in
Eq.~(\ref{propagator}), we see that the spectral function of the 
(pseudo-) NG boson is
\begin{equation}
A(\omega,\vec{k}) = \frac{\omega\Gamma}{(\omega^{2}
- e_{k}^{2}-\Gamma^{2}/4)^{2}+\omega^{2}\Gamma^{2}}\:,
\label{spectral-density}
\end{equation}
where $e_{k} \equiv \sqrt{v^{2}k^2+m^{2}}$. Because of the rotational 
symmetry of the system, the conductivity is characterized by a single 
scalar quantity $\kappa$ which is introduced as follows: $\kappa_{ij} 
=\kappa \delta_{ij}$. The explicit expression for this scalar function 
reads
\begin{equation}
\kappa = \frac{1}{48\sqrt{2} \pi^{2} v \Gamma T^2}  
\int_{0}^{\infty}\frac{\omega d\omega}
{\sinh^{2}\!\frac{\omega}{2T}} 
\left(\sqrt{X^{2}+\omega^{2}\Gamma^{2}}+X\right)^{3/2}\!,
\label{kap-final}
\end{equation}
where we introduced the notation $X\equiv \omega^{2}-m^2-\Gamma^{2}/4$.
For our purposes it will be sufficient to consider the conductivity in
the limit of a small width, $\Gamma\to 0$. This is because the (pseudo-)
NG bosons in the CFL quark matter are weakly interacting. Thus, we derive
the following approximate relation:
\begin{equation}
\kappa = \frac{1}{24 \pi^{2} v T^2\Gamma} \int_{m}^{\infty}
\frac{d\omega \omega}{\sinh^{2}\!\frac{\omega}{2T}}
\left(\omega^{2}-m^2\right)^{3/2}.
\label{kap-G0}
\end{equation}
At small temperature, $T\ll m$, this result is further approximated 
by
\begin{equation}
\kappa \simeq \frac{m^{5/2}\sqrt{T}}
{2\sqrt{2}\pi^{3/2} v \Gamma} e^{-m/T}.
\label{kap-mass}
\end{equation}
This demonstrates clearly that the contributions of heavy 
pseudo-NG bosons to the thermal conductivity are suppressed. 
The largest contribution comes from the massless NG boson $\phi$ 
for which the thermal conductivity is
\begin{equation}
\kappa_{\phi} = \frac{4 T^{3}}{3\pi^{2} v \Gamma_{\phi} } 
\int_{0}^{\infty} \frac{x^{4} dx }{\sinh^{2}\!x}
= \frac{2 \pi^{2}T^{3}}{45 v \Gamma_{\phi} }\:. 
\label{kap-m0}
\end{equation}
In order to calculate $\kappa_{\phi}$ the width $\Gamma_{\phi}$ 
is required, or equivalently the mean free path $\ell_{\phi}$ since
$\ell_{\phi} \equiv\bar{v}/\Gamma_{\phi}$, where $\bar{v}$ is the 
average thermal velocity of the NG bosons. This will be discussed 
in the next section.

\section{Mean free path of the NG boson}
\label{mfp}

As we have remarked, the contribution of massive pseudo-NG bosons to the
thermal conductivity is suppressed.  In the CFL phase of quark matter,
however, there is one truly massless NG boson $\phi$ which should
therefore give the dominant contribution to the heat conductivity. The
interactions of $\phi$ with the CFL matter leads to a finite value for
its mean free path. Since this boson is a composite particle there is
always a non-zero probability at finite temperature for its decay into a
pair of quark quasiparticles. It is natural to expect that such a process
is strongly suppressed at small temperatures, $T\ll\Delta$. This is
confirmed by a direct microscopic calculation in the region of asymptotic
densities which yields a decay width \citebk{GS}:
\begin{equation}
\Gamma_{\phi\to q q} (k) 
\simeq \frac{5\sqrt{2}\pi v k}{4(21-8\ln2)}
\exp\left(-\sqrt{\frac{3}{2}} 
\frac{\Delta}{T}\right) . 
\label{decay-quark}
\end{equation}
If this were the only contribution, then the order of magnitude of
the mean free path of the NG boson would be 
\begin{equation}
\ell_{\phi\to q q} \sim \frac{v}{T} \exp\left(\sqrt{\frac{3}{2}}
\frac{\Delta}{T}\right).
\end{equation}
This grows exponentially with decreasing temperature. For example, if
$\Delta \simeq 50$ MeV and $T\lesssim 1.5$ MeV, the mean free path is 
$30$ km or more. This scale is a few times larger than the typical 
size of a compact star.

The decay channel of the NG bosons into quarks is not the only 
contribution to the mean free path. They can also scatter on one another.
The corresponding amplitude is of order $k^4/\mu^{4}$ \citebk{son}
which gives a cross section of $\sigma_{\phi\phi} \simeq T^{6}/\mu^8$,
yielding the following contribution to the width:
\begin{equation}
\Gamma_{\phi\phi} = v \sigma_{\phi\phi} n_{\phi}
\sim \frac{T^{9}}{\mu^8}\:,
\label{self-int}
\end{equation}
where $n_{\phi}$ is the equilibrium number density of the NG bosons
\citebk{SE-prc}. At small temperatures the scattering contribution 
in Eq.~(\ref{self-int}) dominates the width. This leads to a mean 
free path 
\begin{eqnarray}
\ell_{\phi\phi} \sim \frac{\mu^8}{T^{9}}
\approx 8\times 10^5\frac{\mu_{500}^8}{T_{\rm MeV}^{9}}\mbox{ km}.
\end{eqnarray}
Here we defined the following dimensionless quantities: $\mu_{500} \equiv
\mu/(500\mbox{ MeV})$ and $T_{\rm MeV} \equiv T/(1\mbox{ MeV})$.  Both
$\ell_{\phi\phi}$ and $\ell_{\phi\to qq}$ depend very strongly on
temperature, however the salient point is that they are both larger than
the size of a compact star for temperatures $T_{\rm MeV}$ of order 1.

We define $\tilde{T}$ to be the temperature at which the massive NG
bosons decouple from the system. This is determined by the mass of the
lightest pseudo-NG boson for which it is not presently possible to give a
reliable value. Different model calculations \citebk{SonSt,BS,SchMass}
produce different values which can range as low as 10 MeV. Thus,
conservatively, we choose $\tilde{T}\simeq 1$ MeV. Then the mean free
path of the NG boson is comparable to or even larger than the size of a
star for essentially all temperatures $T\lesssim \tilde{T}$. It is also
important to note that the mean free path is very sensitive to
temperature changes. In particular, at temperatures just a few times
higher than $\tilde{T}$ the value of $\ell$ may already become much
smaller than the star size. This suggests that, during the first few
seconds after the supernova explosion when the temperatures remain
considerably higher than $\tilde{T}$, a noticeable temperature gradient
may exist in the quark core. This should relax very quickly because of
the combined effect of cooling (which is very efficient at $T\gg
\tilde{T}$) and diffusion. After that almost the whole interior of the
star would become isothermal.

Before concluding this section, we point out that the geometrical size of
the quark phase limits the mean free path of the NG boson since the
scattering with the boundary should also be taken into account. It is
clear from simple geometry that $\ell \sim R_{0}$, where $R_{0}$ is the
radius of the quark core.

\section{Photon contributions}
\label{photon}

Now, let us discuss the role of photons in the CFL quark core. It was
argued in Ref.~\citebk{JPS} that the mean free path of photons is larger
than the typical size of a compact star at all temperatures $T\lesssim
\tilde{T}$. One might conclude therefore that all photons would leave the
stellar core shortly after the core becomes transparent. If this were so
the photons would be able to contribute neither to the thermodynamic nor
to the kinetic properties of the quark core. However the neighboring
neutron matter has very good metallic properties due to the presence of a
considerable number of electrons. As is known from plasma physics, low
frequency electromagnetic waves cannot propagate inside a plasma.
Moreover, an incoming electromagnetic wave is reflected from the surface
of such a plasma \citebk{Plasma}.  In particular, if $\Omega_p$ is the
value of the plasma frequency of the nuclear matter, then all photons
with frequencies $\omega < \Omega_p$ are reflected from the boundary.
This effect is similar to the well known reflection of radio waves from
the Earth's ionosphere.

The plasma frequency is known to be proportional to the square root 
of the density of charge carriers and inversely proportional to the
square root of their mass. It is clear therefore that the electrons,
rather than the more massive protons, will lead to 
the largest value of the plasma frequency in nuclear matter.
Our estimate for the value of this frequency is
\begin{equation}
\Omega_p=\sqrt{\frac{4\pi e^2 Y_e \rho}{m_e m_p}}
\simeq 4.7\times 10^2 \sqrt{\frac{\rho Y_e}{\rho_{0}}}\mbox{ MeV}, 
\end{equation}
where the electron density $n_e = Y_e \rho/m_p$ is given in 
terms of the nuclear matter density $\rho$ and the proton mass
$m_p$. Also $m_e$ denotes the electron mass, 
$Y_e\simeq 0.1$ is the number of electrons per baryon, 
and $\rho_{0}\approx 2.8 \times 10^{14} \mbox{ g cm}^{-3}$ 
is equilibrium nuclear matter density.

Since $\Omega_p$ is more than $100$ MeV, essentially all thermally 
populated electromagnetic waves at $T\lesssim \tilde{T}$ will be 
reflected back into the core region. In a way the boundary of the core 
looks like a good quality mirror with some leakage which will allow 
a thermal photon distribution to build up and stay. Thus photons will 
be {\it trapped} in such a core surrounded by a nuclear layer. Notice 
that the transparency of the core is reached only after the temperature 
drops substantially below $\tilde{T}\simeq 2m_{e}$, i.e., when the 
density of thermally excited electron-positron pairs becomes very small. 

Now, since photons are massless they also give a sizable contribution
to the thermal conductivity of the CFL phase. The corresponding 
contribution $\kappa_{\gamma}$ will be similar to the contribution of 
massless NG bosons in Eq.~(\ref{kap-m0}). Since the photons move
at approximately the speed of light \citebk{Litim:2001mv} ($v\simeq1$ at 
the densities of interest) and they have two polarization states, we obtain
\begin{equation}
\kappa_{\gamma} =\frac{4 \pi^{2}T^{3}}{45 \Gamma_{\gamma}}\:.
\end{equation}

Since the thermal conductivity is additive the total conductivity of
dense quark matter in the CFL phase is given by the  sum of the two 
contributions:
\begin{equation}
\kappa_{CFL} =\kappa_{\phi} +\kappa_{\gamma}
\simeq \frac{2\pi^2}{9} T^{3} R_{0},
\end{equation}
where for both a photon and a NG boson the mean free path
$\ell\sim R_0$.
This yields the value
\begin{equation}
\kappa_{CFL} \simeq 1.2 \times 10^{32} T_{\rm MeV}^{3} R_{0,{\rm km}}
\mbox{ erg cm}^{-1} \mbox{sec}^{-1} \mbox{K}^{-1},
\end{equation}
where $R_{0,{\rm km}}$ is the quark core radius measured in kilometers. 
The value of $\kappa_{CFL}$ is many orders of magnitude larger than the 
thermal conductivity of regular nuclear matter in a neutron star 
\citebk{van}.

\section{Stellar cooling}
\label{cool}

In discussing the cooling mechanism for a compact star we have to make
some general assumptions about the structure of the star. We accept
without proof that a quark core exists at the center of the star. This
core stays in direct contact with the neighboring nuclear matter. From
this nuclear layer outwards the structure of the star is essentially the
same as an ordinary neutron star. The radius of the core is denoted by
$R_{0}$, while the radius of the whole star is denoted by $R$.

A detailed analysis of the interface between the quark core and the
nuclear matter was made in Ref.~\citebk{Interface}. A similar analysis
might also be very useful for understanding the mechanism of heat
transfer from one phase to the other. Here we assume that direct contact
between the phases is possible, and that the temperature is slowly
varying across the interface.

Now, let us consider the physics that governs stellar cooling. We start
from the moment when the star is formed in a supernova explosion.
Immediately after the explosion many high-energy neutrinos are trapped
inside the star. After about $10$ to $15$ seconds most of them escape
from the star by diffusion. The presence of the CFL quark core could
slightly modify the rate of such diffusion
\citebk{CarRed,Nu-diffusion,0203011}. By the end of the deleptonization
process, the temperature of the star will have risen to a few tens of
MeV. Then, the star cools down relatively quickly to about $\tilde{T}$ by
the efficient process of neutrino emission.  It is unlikely that the
quark core would greatly affect the time scale for this initial cooling
stage. An ordinary neutron star would continue to cool by neutrino
emission for quite a long time even after that \citebk{PLPSR}. Here we
discuss how the presence of the CFL quark core affects the cooling
process of the star after the temperature drops below $\tilde{T}$.

Our result for the mean free path of the NG boson demonstrates that the
heat conductivity of dense quark matter in the CFL phase is very high.
For example, a temperature gradient of $1$ MeV across a core of $1$ km in
size is washed away by heat conduction in a very short time interval of
order $R_{0,{\rm km}}^2/v\ell(T) \simeq 6\times 10^{-4}$ sec. In deriving
this estimate, we took into account the fact that the specific heat and
the heat conductivity in the CFL phase are dominated by photons and
massless NG bosons and that $\ell\sim R_0$. In addition, we used the
classical relation $\kappa = \bar{v} c_{v} \ell /3$, where $c_{v}$ is the
specific heat; this can be shown to hold in the present context
\citebk{SE-prc}. Since heat conduction removes a temperature gradient in
such a short time interval, it is clear that, to a good approximation,
the quark core is isothermal at all temperatures $T\lesssim\tilde{T}$.

The heat conductivity of the neighboring nuclear matter is also known to
be very high because of the large contribution from degenerate electrons
which have a very long mean free path. It is clear, then, that both the
quark and the nuclear layers should be nearly isothermal with equal
values of the temperature. When one of the layers cools down by any
mechanism, the temperature of the other will adjust almost immediately
due to the very efficient heat transfer on both sides of the interface.

Now consider the order of magnitude of the cooling time for a star with a
CFL quark core. One of the most important components of the calculation
of the cooling time is the thermal energy of the star which is the amount
of energy that is lost in cooling.  There are contributions to the total
thermal energy from both the quark and the nuclear parts of the star. The
dominant amount of thermal energy in the CFL quark matter is stored in
photons and massless NG bosons. Numerically, its value is \citebk{SE-prc}
\begin{equation}
E_{CFL}(T) \simeq 2.1 \times 10^{42} R_{0,{\rm km}}^{3} 
T_{\rm MeV}^{4} \mbox{ erg}.
\end{equation}
The thermal energy of the outer nuclear layer is provided mostly by 
degenerate neutrons. The corresponding numerical estimate is 
\citebk{Shapiro}:
\begin{equation}
E_{NM}(T) \simeq 8.1 \times 10^{49} \frac{M-M_{0}}{M_{\odot}}
\left(\frac{\rho_{0}}{\rho}\right)^{2/3} T^{2}_{\rm MeV}
\mbox{ erg}, 
\end{equation}
where $M$ is the mass of the star, $M_{0}$ is the mass of the quark core
and $M_{\odot}$ is the mass of the Sun.  It is crucial to note that the
thermal energy of the quark core is negligible in comparison to that of
the nuclear layer.

The second important component that determines stellar cooling is the
luminosity which describes the rate of energy loss due to neutrino and
photon emission. Typically, the neutrino luminosity dominates the cooling
of young stars when the temperatures are still higher than about $10$ keV
and after that the photon diffusion mechanism starts to dominate. Photon
and neutrino emission from the CFL quark phase is strongly suppressed at
low temperatures \citebk{JPS}. The neighboring nuclear layer, on the
other hand, emits neutrinos quite efficiently. As a result, it cools
relatively fast in the same way as an ordinary neutron star. The nuclear
layer should be able to emit not only its own thermal energy, but also
that of the quark core which constantly arrives by the very efficient
heat transfer process. The analysis of the cooling mechanism is greatly
simplified by the fact that the thermal energy of the quark core is
negligible compared to the energy stored in the nuclear matter. By taking
this into account, we conclude that the cooling time of a star with a
quark core is essentially the same as for an ordinary neutron star
provided that the nuclear layer is not extremely thin.

\section{Conclusions}
\label{conclusion}

Our analysis shows that the thermal conductivity of CFL color
superconducting dense quark matter is very high for typical values of the
temperature found in a newborn compact star. This is a direct consequence
of the existence of the photon and the massless NG boson $\phi$ whose
mean free paths are very large. Note that the photons are trapped in the
core because of reflection by the electron plasma in the neighboring
nuclear matter. The NG bosons are also confined to the core since they
can only exist in the CFL phase.

It is appropriate to mention that the (pseudo-) NG bosons and photons
should also dominate other kinetic properties of dense quark matter in
the CFL phase. For example, the shear viscosity should be mostly due to
photons and the same massless NG bosons associated with the breaking of
baryon number. The electrical conductivity, on the other hand, would be
mostly due to the lightest {\em charged} pseudo-NG boson, i.e., the
$K^{+}$. Thus, in the limit of small temperatures, $T\to 0$, the
electrical conductivity will be suppressed by a factor
$\exp(-m_{K^{+}}/T)$.

Since the neutrino emissivity of the CFL core is strongly suppressed, the
heat is transferred to the outer nuclear layer only through direct
surface contact. While both the core and the outer layer contribute to
the heat capacity of the star, it is only the outer layer which is
capable of emitting this heat energy efficiently in the form of
neutrinos. The combination of these two factors tends to extend the
cooling time of a star. However, because of the very small thermal energy
of the quark core, the time scale for cooling could be noticeably longer
than that for an ordinary neutron star only if the outer nuclear 
layer was very thin. (Note that, while little is known about the 
properties of thin boundary layers outside CFL matter, it is possible that
photon emission from this layer might quickly drain the relatively 
small amount of CFL thermal energy.)
Thus it appears that the cooling of stars with
not too large CFL quark cores will differ little from the cooling of
typical neutron stars. A similar conclusion has been reached for stars with regular, 
non-CFL quark interiors \citebk{ppls}.

In passing it is interesting to speculate about the possibility that a
bare CFL quark star made entirely of dense quark matter could exist. If
it were possible, it would look like a transparent dielectric
\citebk{RW-diamond}. Our present study suggests such a star would also
have very unusual thermal properties. Indeed, if the star has a finite
temperature $T\lesssim \tilde{T}$ after it was created, almost all of its
thermal energy would be stored in the NG bosons. Notice that all the
photons would leave the star very soon after transparency set in because
the star is assumed to have no nuclear matter layer. The local
interaction as well as the self-interaction of the NG bosons is very weak
so that we argued in Sec.~\ref{mfp} that their mean free path would be
limited only by the geometrical size of the star. This suggests that,
since photon and neutrino emission inside the CFL phase is strongly
suppressed \citebk{JPS}, the
only potential source of energy loss in the bare CFL star would be the
interaction of the NG bosons at the stellar boundary and 
photon emission there. It is likely, therefore, that such stars might
be very dim and might even be good candidates for some of the baryonic
dark matter in the Universe.

\section*{Acknowledgments}
This work was supported by the U.S. Department of Energy Grant
No.~DE-FG02-87ER40328.


\begin{thebibliography}{0}

\bibitem{ColPer} J.C.~Collins and M.J.~Perry,
{\it Phys. Rev. Lett.} {\bf 34}, 1353 (1975).

\bibitem{old} B.~C.~Barrois, 
{\it Nucl. Phys. } {\bf B129}, 390 (1977);
S.~C.~Frautschi,
in {\it Hadronic Matter at Extreme Energy Density}, edited by 
N.~Cabibbo and L.~Sertorio (Plenum, New York, 1980);
D.~Bailin and A.~Love,
{\it Phys. Rep.}  {\bf 107}, 325 (1984).

\bibitem{W1S1} M.~G.~Alford, K.~Rajagopal and F.~Wilczek,
{\it Phys. Lett.}  {\bf B422}, 247 (1998);
R.~Rapp, T.~Sch\"{a}fer, E.~V.~Shuryak and M.~Velkovsky,
{\it Phys. Rev. Lett.}  {\bf 81}, 53 (1998).
 
\bibitem{PR1-Son} D.~T.~Son,
{\it Phys. Rev.}  {\bf D59}, 094019 (1999);
R.~D.~Pisarski and D.~H.~Rischke,
{\it Phys. Rev. Lett.}  {\bf 83}, 37 (1999).

\bibitem{2nd-wave} T.~Schafer and F.~Wilczek,
{\it Phys. Rev.}  {\bf D60}, 114033 (1999);
D.~K.~Hong, V.~A.~Miransky, I.~A.~Shovkovy and L.~C.~R.~Wijewardhana,
{\em ibid.} {\bf D61}, 056001 (2000);
{\bf D62}, 059903(E) (2000);
R.~D.~Pisarski and D.~H.~Rischke,
{\em ibid.} {\bf D61}, 051501 (2000).

\bibitem{3rd-wave} S.~D.~Hsu and M.~Schwetz,
{\it Nucl. Phys.}  {\bf B572}, 211 (2000);
W.~E.~Brown, J.~T.~Liu and H.~C.~Ren,
{\it Phys. Rev.}  {\bf D61}, 114012 (2000).

\bibitem{us2} I.~A.~Shovkovy and L.~C.~R.~Wijewardhana,
{\it Phys. Lett.}  {\bf B470}, 189 (1999);
T.~Sch\"{a}fer,
{\it Nucl. Phys.} {\bf B575}, 269 (2000).

\bibitem{ARW} M.~Alford, K.~Rajagopal and F.~Wilczek,
{\it Nucl. Phys.}  {\bf B537}, 443 (1999);
note that the 2SC phase involving just two quark flavors is 
thought to be absent in compact stars, see
M.~Alford and K.~Rajagopal, {\it JHEP} {\bf 0206}, 031 (2002). 

\bibitem{CasGat} R.~Casalbuoni and R.~Gatto,
{\it Phys. Lett.} {\bf B464}, 111 (1999).

\bibitem{SonSt} D.~T.~Son and M.~A.~Stephanov,
{\it Phys. Rev.}  {\bf D61}, 074012 (2000);
{\bf D62}, 059902(E) (2000).

\bibitem{BS} P.~F.~Bedaque and T.~Sch\"{a}fer,
{\it Nucl. Phys.}  {\bf A697}, 802 (2002);
D. B. Kaplan and S. Reddy, 
{\it Phys. Rev.} {\bf D65}, 054042 (2002).

\bibitem{MirSho} V.~A.~Miransky and I.~A.~Shovkovy,
{\it Phys. Rev. Lett.} {\bf 88}, 111601 (2002);
T.~Schafer, D.~T.~Son, M.~A.~Stephanov, D.~Toublan and J.~J.~Verbaarschot,
{\it Phys. Lett.}  {\bf B522}, 67 (2001).

\bibitem{JPS} P.~Jaikumar, M.~Prakash and T.~Sch\"{a}fer,
astro-ph/0203088.

\bibitem{SE-prc} I.~A.~Shovkovy and P.~J.~Ellis,
{\it Phys. Rev.} {\bf C66}, 015802 (2002).

\bibitem{fgi} E. J. Ferrer, V. P. Gusynin and V. de la Incera,
cond-matt/0203217.

\bibitem{GS} V.~P.~Gusynin and I.~A.~Shovkovy,
{\it Nucl. Phys.}  {\bf A700}, 577 (2002).

\bibitem{son} D. T. Son, hep-ph/0204199.

\bibitem{SchMass} T.~Sch\"{a}fer,
hep-ph/0201189.

\bibitem{Plasma} P.~A.~Sturrock, {\it Plasma Physics}
(Cambridge University Press, Cambridge, 1994).

\bibitem{Litim:2001mv} D.~F.~Litim and C.~Manuel,
{\it Phys. Rev.} {\bf D64}, 094013 (2001).

\bibitem{van} J. M. Lattimer, K. A. Van Riper, M. Prakash and M. Prakash,
{\it Astrophys. J.}  {\bf 425}, 802 (1994).
 
\bibitem{Interface} M.~G.~Alford, K.~Rajagopal, S.~Reddy and F.~Wilczek,
{\it Phys. Rev.} {\bf D64}, 074017 (2001).

\bibitem{CarRed}
G.~W.~Carter and S.~Reddy,
{\it Phys. Rev.} {\bf D62}, 103002 (2000).

\bibitem{Nu-diffusion} A.~W.~Steiner, M.~Prakash and J.~M.~Lattimer,
{\it Phys. Lett.} {\bf B509}, 10 (2001).

\bibitem{0203011} S.~Reddy, M.~Sadzikowski and M.~Tachibana,
nucl-th/0203011.

\bibitem{PLPSR} 
M.~Prakash, J.~M.~Lattimer, J.~A.~Pons, A.~W.~Steiner and S.~Reddy,
{\it Lect. Notes Phys.}  {\bf 578}, 364 (2001).

\bibitem{Shapiro} S.~L.~Shapiro and S.~A.~Teukolsky, {\it Black
Holes, White Dwarfs, and Neutron Stars: The Physics of Compact
Objects} (Wiley, New York, 1983).

\bibitem{ppls} D. Page, M. Prakash, J. M. Lattimer and A. W. Steiner,
{\it Phys. Rev. Lett.} {\bf 85}, 2048 (2000).

\bibitem{RW-diamond} K.~Rajagopal and F.~Wilczek,
{\it Phys. Rev. Lett.}  {\bf 86}, 3492 (2001).

\end{thebibliography}
\end{document}